\documentclass[a4paper,12pt]{article}

\usepackage[left=2.5cm,right=2.5cm,top=2.5cm,bottom=2.5cm]{geometry}
\usepackage{graphicx}
\usepackage{amssymb}
\usepackage{multirow}
\usepackage{amsmath}
\usepackage{psfrag}
\usepackage{setspace}
\usepackage{subcaption}
\usepackage[colorlinks=true,bookmarks=true]{hyperref}
\usepackage{float}
\usepackage{cite}
\usepackage[utf8]{inputenc}

\hypersetup{citecolor=blue}
\newcommand{\dif}{\mathrm{d}}
\newcommand{\Eqref}[1]{(\ref{#1})}
\newcommand{\half}{\frac{1}{2}}

\newcommand{\brac}[1]{\left(#1 \right)}
\newcommand{\sbrac}[1]{\left[#1\right]}
\newcommand{\im}{\mathrm{i}}

\newcommand{\Pcal}{\mathcal{P}}
\newcommand{\Qcal}{\mathcal{Q}}

\begin{document}

\title{Electric/magnetic universe with a cosmological constant}
\author{Yen-Kheng Lim\footnote{E-mail: yenkheng.lim@gmail.com}\\\textit{\normalsize{Department of Mathematics, Xiamen University Malaysia,}}\\\textit{\normalsize{43900 Sepang, Malaysia}}}
\date{\normalsize{\today}}
\maketitle

\begin{abstract}
  A spacetime consisting of parallel electric/magnetic fields held together by its own gravity in the presence of a cosmological constant $\Lambda$ is derived as a limit of the de Sitter/anti-de Sitter C-metric. The limiting procedure is similar to the $\Lambda=0$ case where the Melvin universe is derived from the C-metric. Under an appropriate coordinate transformation, we show that this solution is equivalent to the solution obtained by Astorino. Some physical and geometrical properties of the solution are studied, as well as its geodesics. For $\Lambda<0$, the solution is asymptotically locally anti-de Sitter, and can be derived as a double-Wick rotation of a charged anti-de Sitter black hole with a planar horizon.
\end{abstract}

\section{Introduction} \label{intro}

The Melvin universe \cite{Bonnor:1954,Melvin:1963qx,Melvin:1965zza} is an electro-vacuum solution to the Einstein-Maxwell equations with a vanishing cosmological constant. Physically, the metric describes a set of parallel electric/magnetic fields held together by its own gravity. The study of these spacetimes was initially motivated by issues related to the gravitational stability of electro-magnetic configurations \cite{PhysRev.139.B244}, as well as the study of electro-magnetic `geons' \cite{Wheeler:1955}.\footnote{For more references and a wider overview, see, e.g., \cite{Griffiths:2009dfa}.} Subsequently, magnetic spacetimes came to more observational relevance as strong magnetic fields are believed to be the mechanism behind various astrophysical and cosmological phenomena \cite{Thompson:1995gw,Archibald:2013kla,Durrer:2013pga}. Specifically, Refs.~\cite{Bicak:1985,Kardashev:1995} investigated these phenomena using the framework of a spacetime depicting a magnetised black hole.

Aside from directly solving the Einstein-Maxwell equations, one could derive the Melvin solution by magnetising the Minkowski spacetime. The electric solution is obtained by taking the electro-magnetic dual of the result. The magnetisation procedure is achieved by applying the Harrison transformation \cite{Harrison:1968} on a seed metric. This transformation exploits a symmetry in the Einstein-Maxwell equations with vanishing cosmological constant and can be applied to various axi-symmetric spacetimes, such as black holes and the C-metric.

The second method is of more relevance to this paper, namely by extracting the solution from the charged C-metric \cite{Ehlers:1962,Kinnersley:1970zw,Hong:2003gx,Griffiths:2006tk}. The charged C-metric describes a pair of causally-disconnected, charged black holes uniformly accelerating apart. The intuition behind this limiting procedure is straightforward: The black holes are pushed far away, while simultaneously increasing their charges, one expects that the electro-magnetic fields in the region at the centre of the two black holes will tend to a homogeneous configuration. Indeed, it was demonstrated by Havrdov\'{a} and Krtou\v{s} that the resulting limit is precisely the Melvin spacetime \cite{Havrdova:2006gi}. Interestingly, a similar procedure was performed by Bi\v{c}\'{a}k et al. \cite{Bicak:1983} on two pairs of uniformly accelerated particles by pushing them apart while simultaneously increasing their \emph{mass}. 

With the discovery of the accelerating expansion of the universe and interest in the AdS/CFT correspondence, recent research in General Relativity is being increasingly favoured towards solutions involving a non-zero cosmological constant $\Lambda$. Therefore it is natural to ask whether there is an analogue to the Melvin solution with a cosmological constant. However, many solution-generating methods are not applicable to Einstein-Maxwell equations with $\Lambda\neq 0$, or that many of its useful symmetries are broken by the presence of $\Lambda$. Nevertheless, Astorino \cite{Astorino:2012zm} managed to obtain the Melvin universe with a cosmological constant by an appropriate generalisation of the Ernst potential formalism \cite{Ernst:1967wx,Ernst:1967by}. More recently, the Melvin universe with $\Lambda\neq0$ has been considered in the context of gyration solutions \cite{Kadlecova:2016irj}, where the spacetime additionally includes spinning, ultra-relativistic matter sources \cite{Kadlecova:2009qu,Kadlecova:2010je,Kadlecova:2013uta}.

In this paper, we derive the electric/magnetic Melvin universe with a non-zero cosmological constant by by applying Havrdov\'{a} and Krtou\v{s}'s procedure to the charged \emph{de Sitter/anti-de Sitter C-metric} \cite{Dias:2002mi,Dias:2003xp,Krtous:2005ej,Podolsky:2000pp,Podolsky:2002nk} (henceforth referred to as the (A)dS C-metric). This solution describes a pair of charged black holes accelerating apart in a background de Sitter (dS) or anti-de Sitter spacetime (AdS).\footnote{By `background' dS/AdS spacetime, we mean that the zero mass and charge limit of the solution reduces to that of pure de Sitter or anti-de Sitter spacetime, depending on the sign of the cosmological constant.} We note that Havrdov\'{a} and Krtou\v{s}'s method was made possible due to the fact that the charged C-metric was cast in a factorised form. In particular, the $g_{tt}$ component of the metric takes the form 
\begin{align}
 g_{tt}\propto \brac{\tilde{y}^2-1}\brac{1-2ma\tilde{y}+e^2a^2y^2},
\end{align}
where $a$ and $m$ are the acceleration and mass parameters respectively. Loosely speaking, $\tilde{y}$ is regarded as the inverse radial coordinate, and the horizons of the C-metric are given simply as the roots of the above equation. Particularly, in the notation of \cite{Havrdova:2006gi}\footnote{Note that this is will be different to the notation for the C-metric used in \cite{Hong:2003gx} and the present paper, where $y=-\tilde{y}$ below.} $y=-1$ is the black-hole horizon and $y=+1$ is the acceleration horizon.  Hence a limiting procedure can be expressed in coordinates with a simple form $y=1+\mathcal{O}\brac{\epsilon}$. Recently, the factorised form of the (A)dS C-metric was given in \cite{Chen:2015vma}, thus allowing the possibility to apply the same intuition of pushing the black holes far away while simultaneously increasing the charge. For convenience, we will denote the result as the \emph{dS Melvin} or \emph{AdS Melvin} spacetimes. (Collectively, the \emph{(A)dS Melvin} spacetime.) As will be shown below, this solution is an electro-vacuum spacetime consisting of parallel electric/magnetic fields, though the metric may no longer be cylindrically symmetric like the original Melvin universe. 

This paper is organised as follows. In Sec.~\ref{Derivation}, review some relevant features of the (A)dS C-metric before taking the limit to the (A)dS Melvin solution. For the case of the negative cosmological constant, we show that the AdS magnetic Melvin solution can also be derived from the charged version of the AdS soliton. Some physical and geometrical properties of the (A)dS Melvin solution are explored in Sec.~\ref{Physical}. In Sec.~\ref{OtherForms}, we show that our (A)dS Melvin solution is equivalent to Astorino's by an appropriate coordinate transformation. We also show how the familiar $\Lambda=0$ Melvin solution can be recovered, along with other interesting limiting cases. Subsequently, the motion of test particles in this solution is explored in Sec.~\ref{Geodesics}. We briefly comment on the possibility of generalisation to higher dimensions in Sec.~\ref{HigherD}. This paper concludes with a brief discussion in Sec.~\ref{Discussion}.

\section{Derivation of the (A)dS Melvin spacetime} \label{Derivation}

\subsection{Review of the (A)dS C-metric}

Let us establish the notation to be used in this paper and briefly review some essential features of the (A)dS C-metric that will be relevant for the derivation of the (A)dS Melvin spacetime. Further details of the metric have been studied in Ref.~\cite{Chen:2015vma}. The (A)dS C-metric, along with the Melvin limit and other spacetimes discussed in this paper\footnote{With the exception of Sec.~\ref{HigherD}, which discusses the Einstein-Maxwell equations in higher dimensions.} are solutions to the four-dimensional Einstein-Maxwell equations
\begin{subequations} \label{EME}
\begin{align}
 R_{\mu\nu}&=\Lambda g_{\mu\nu}+2F_{\mu\lambda}{F_\nu}^\lambda-\half F^2g_{\mu\nu},\label{EE}\\
 \nabla_\lambda F^{\lambda\nu}&=0,\label{ME}
\end{align}
\end{subequations}
where $F_{\mu\nu}$ is the Maxwell tensor derived from the gauge potential $A$ by $F=\dif A$. We have also denoted $F^2=F_{\mu\nu}F^{\mu\nu}$. In four dimensions, we shall use the (A)dS scale $\ell$ which is related to the cosmological constant by
\begin{align}
 \Lambda=-\frac{3}{\ell^2}.
\end{align}
Therefore, solutions with a positive cosmological constant is understood to have $\ell^2<0$ and those with a negative cosmological constant accordingly has $\ell^2>0$.

The (A)dS C-metric solution is given by \cite{Chen:2015vma}
\begin{subequations}
\begin{align}
\label{AdSCmet}
\begin{split}
 \dif s^2&=\frac{-\ell^2(a^2-1)(b^2-1)}{(x-y)^2}\brac{-\Qcal(y)\dif t^2+\frac{\dif y^2}{\Qcal(y)}+\frac{\dif x^2}{\Pcal(x)}+\Pcal(x)\dif\phi^2},\\
 \Pcal(x)&=\brac{1-x^2}\sbrac{q^2(x-a)(x-b)-(a+b)\brac{x-a-b}-ab-1},\\
 \Qcal(y)&=(y-a)(y-b)\sbrac{q^2(y^2-1)-(a+b)y-ab-1},\\
\end{split}\\
 A&=\sqrt{-\ell^2(a^2-1)(b^2-1)}\brac{ey\,\dif t - gx\,\dif\phi}, \label{AdSCpot}
\end{align}
\end{subequations}
where $e$ and $g$ are the respectively electric and magnetic charges satisfying $e^2+g^2=q^2$. We will restrict our parameter ranges such that $-\ell^2(a^2-1)(b^2-1)>0$, so that the metric will carry a Lorentzian signature $(-+++)$ in its static patch. 

\begin{figure}
 \begin{center}
  \begin{subfigure}[b]{0.49\textwidth}
   \centering
   \includegraphics[scale=0.45]{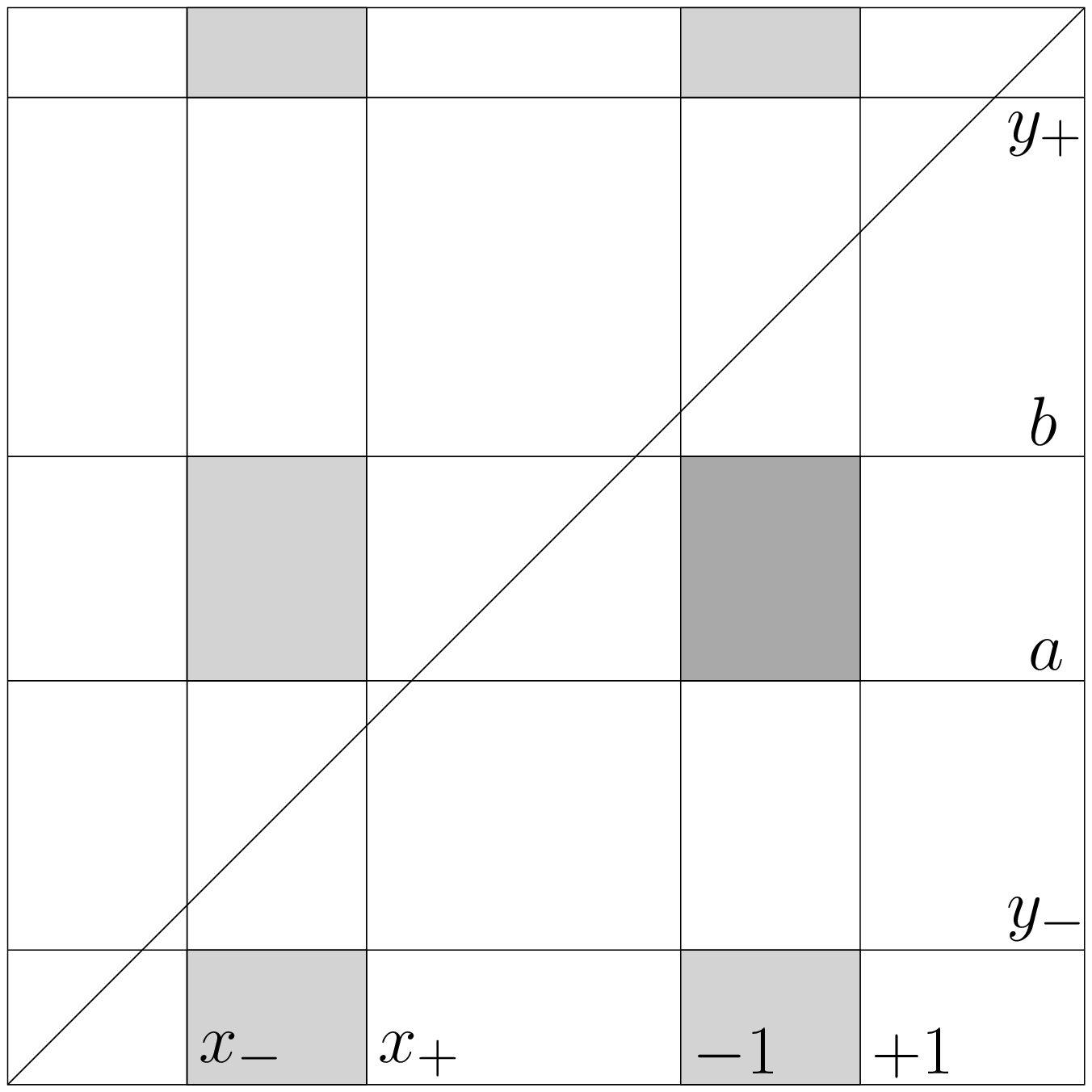}
   \caption{$\ell^2<0$ (dS)}
   \label{fig_DomaindS}
  \end{subfigure}
    \begin{subfigure}[b]{0.49\textwidth}
   \centering
   \includegraphics[scale=0.45]{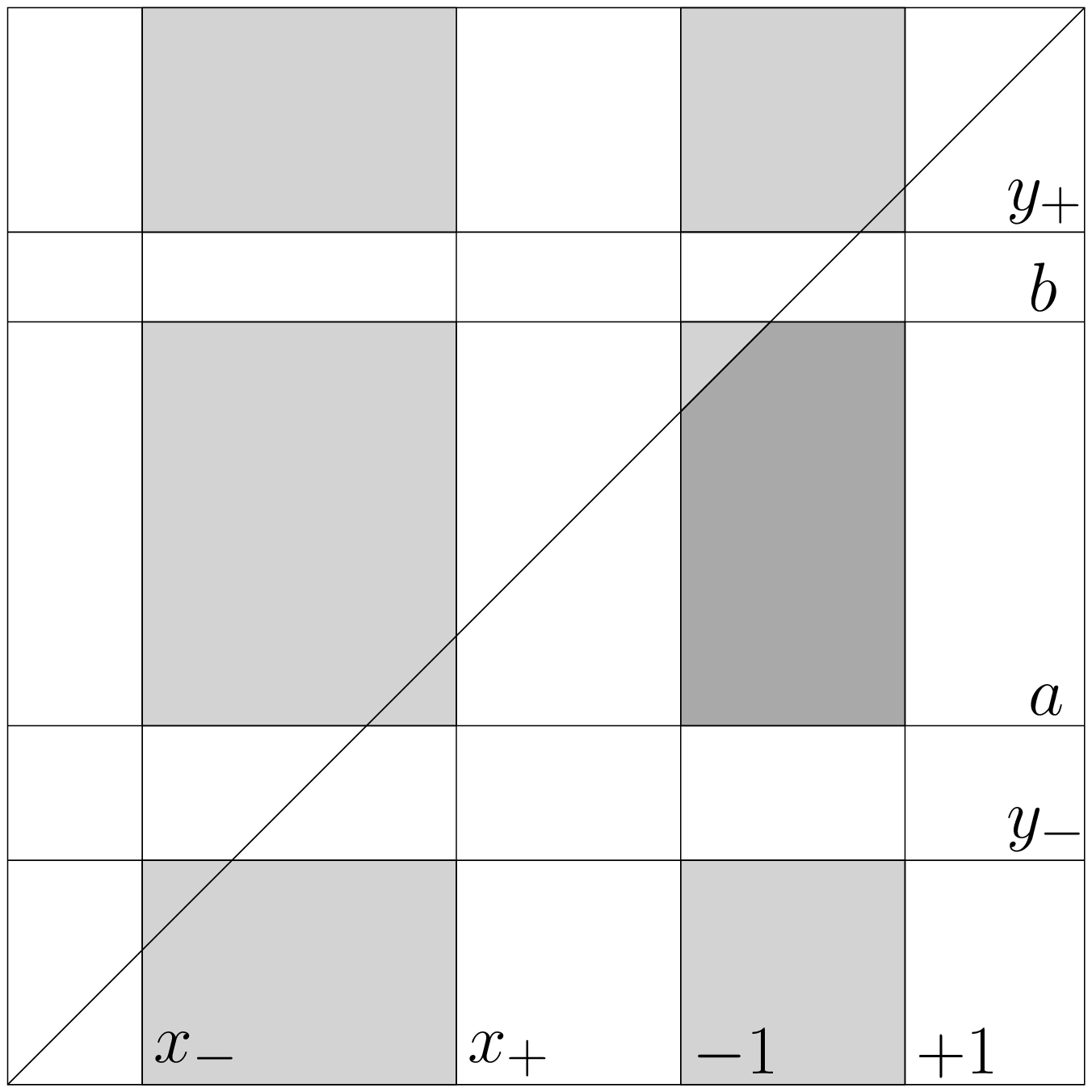}
   \caption{$\ell^2>0$ (AdS)}
   \label{fig_DomainAdS}
   \end{subfigure}\vspace{6pt}
  \begin{subfigure}[b]{0.49\textwidth}
   \centering
   \includegraphics[scale=0.45]{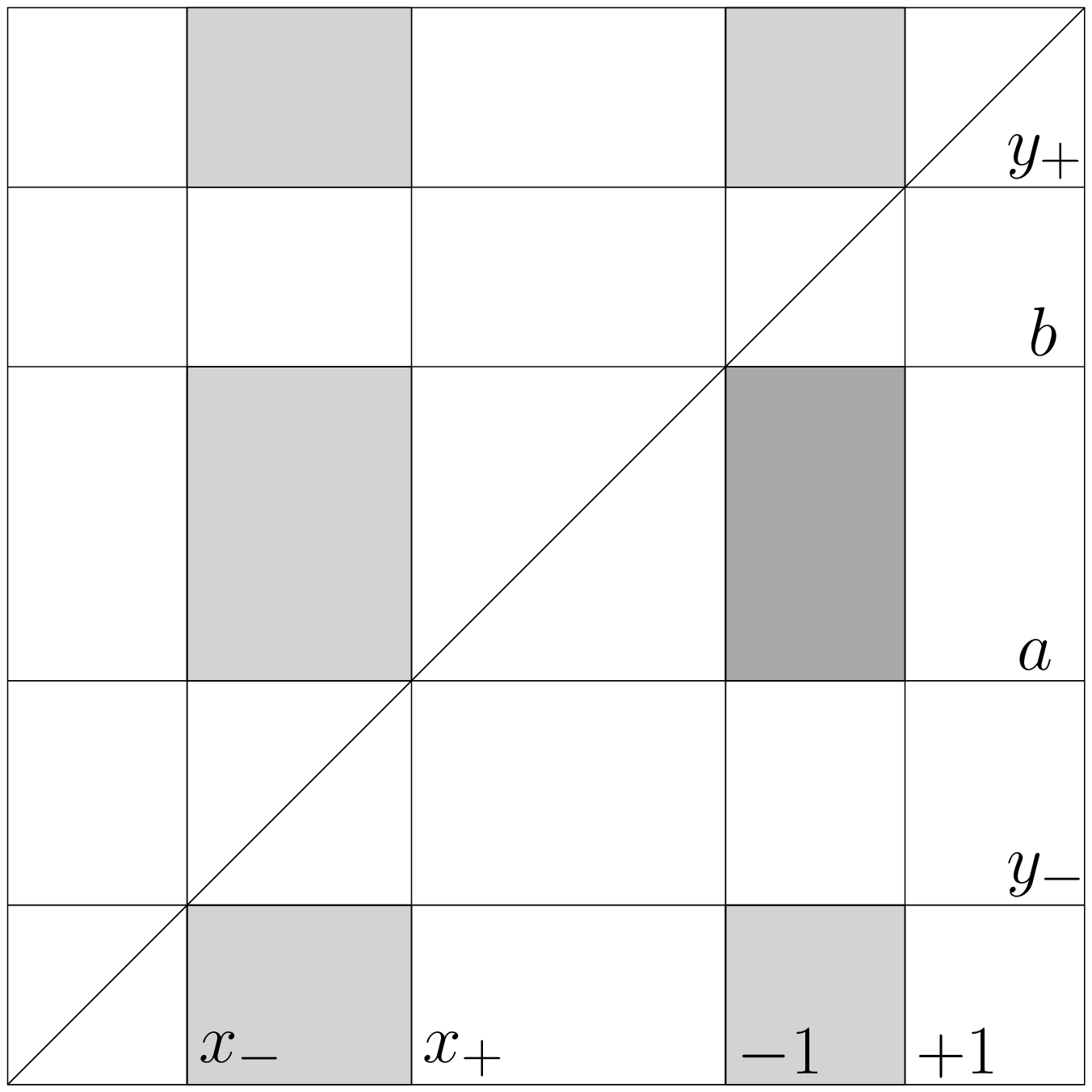}
   \caption{$\ell^2\rightarrow\pm\infty$ (zero cosmological constant)}
   \label{fig_DomainFlat}
  \end{subfigure}
 \end{center}
 \caption[Domains of the charged C-metric]{The Lorentzian domains of the charged C-metric for the cases (a) $\ell^2<0$ (dS), (b) $\ell^2>0$ (AdS), and (c) $\ell^2\rightarrow\pm\infty$ (zero cosmological constant). In each diagram, the $x$-axis is in the horizontal direction, while the $y$-axis is in the vertical direction. The boxes we are interested in are in the darker shade. The diagonal lines correspond to the conformal infinity $x=y$.}
 \label{fig_Domains}
\end{figure}

Some of the important physical properties of the metric are encoded in the roots of the structure functions $\Pcal$ and $\Qcal$,
\begin{align}
 \label{roots}
 \begin{split}
 \Pcal(x)=0:&\quad x=\pm 1,\quad\mbox{and}\quad x_\pm,\\
 \Qcal(y)=0:&\quad y=a,b,\quad\mbox{and}\quad y_\pm,
 \end{split}
\end{align}
where
\begin{subequations}\label{PQroots}
\begin{align}
 x_\pm&=\frac{1}{2q^2}\sbrac{(1+q^2)(a+b)\pm\sqrt{q^4(a-b)^2-2q^2(a^2+b^2-2)+(a+b)^2}},\label{xpmroots}\\
 y_\pm&=\frac{1}{2q^2}\sbrac{(a+b)\pm\sqrt{4q^4+4q^2(1+ab)+(a+b)^2}}.\label{ypmroots}
\end{align}
\end{subequations}
By design, the metric is cast in the form \Eqref{AdSCmet} so that (some of) the roots of the structure functions take simple forms, namely $a$, $b$, and $\pm1$ in \Eqref{roots}. The ordering of the roots determine the global properties of the spacetime, and they depend on the sign of the cosmological constant:
\begin{itemize}
 \item Positive cosmological constant, $\Lambda>0$ (or $\ell^2<0$):
  \begin{align}
   y_-<x_-<x_+<a<b<-1<+1<y_+. \label{dSroots}
  \end{align}
  
 \item Negative cosmological constant, $\Lambda<0$ (or $\ell^2>0$):
  \begin{align}
   x_-<y_-<a<x_+<-1<b<y_+<+1. \label{AdSroots}
  \end{align}
  
 \item Zero cosmological constant, $\Lambda=0$ (or $\ell^2=\pm\infty$):
  \begin{align}
   y_-=x_-<x_+=a<b=-1<+1=y_+.
  \end{align}
\end{itemize}
These three cases are summarised in Fig.~\ref{fig_Domains}, where the shaded regions indicate coordinates where the metric carries a Lorentzian signature $(-+++)$. The darker shades are regions of primary interest, where in addition to having Lorentzian signature with $t$ being time-like, the coordinates lie within the range 
\begin{align}
 -1<x<1,\quad a<y<b,\quad y-x<0, \label{Cmetric_domain}
\end{align}
where $x=y$ indicates the conformal infinity of the metric.

Having established the root orderings of $\mathcal{P}$ and $\mathcal{Q}$, we have a more concrete interpretation of the coordinate boundaries of Eq.~\Eqref{Cmetric_domain}. The black-hole and acceleration horizons are located at $y=a$ and $y=b$, respectively. In the dS case, Fig.~\ref{fig_DomaindS} tells us that both horizons are compact, starting from $x=-1$ and terminating at $x=+1$, where $x$ is interpreted as the sine of a polar angle. In particular, the compact acceleration horizon is reminiscent of an observer in pure de Sitter space perceiving a cosmological horizon in all directions, and that the horizon area is obtained by integrating over the polar and azimuthal angles to give a finite result. On the other hand, in the AdS case, the acceleration horizon has infinite area, as can be seen in Fig.~\ref{fig_DomainAdS} where the $y=b$ line intersects with the conformal infinity $x=y$. Loosely speaking, we say that the acceleration horizon `extends to infinity'. However, the black-hole horizon is still finite, reflecting its $S^2$ horizon topology.\footnote{More precisely, the black-hole horizon has the shape of a deformed sphere. For details, see Ref.~\cite{Chen:2015zoa}.}

\subsection{The limit to the (A)dS-Melvin universe}

To obtain the (A)dS Melvin spacetime, we apply the method of Havrdov\'{a} and Krtou\v{s} by staying within a region close to the acceleration horizon and push the two charged black holes away, while simultaneously increasing the charge. This is achieved by scaling up an infinitesimal neighbourhood close to $y=b$.

Towards this end, we reparametrise the roots and coordinates by
\begin{align}
 \label{reparametrise}
 \begin{split}
 a&=b+\xi_1\epsilon,\quad \quad y_-=b+\xi_2\epsilon,\quad y=b+\xi_1\xi_2\epsilon^2q^2(b-y_+)u,\quad t=\frac{2\eta}{\xi_1\xi_2\epsilon^2q^2H^2(b-y_+)}.
 \end{split}
\end{align}
First, let us see what happens to $q$ under this limit. Comparing the expressions of $y_-$ from \Eqref{reparametrise} and \Eqref{ypmroots}, in the limit $\epsilon\rightarrow 0$, we have
\begin{align}
 e^2+g^2=q^2=\frac{3b^2+1}{b^2-1}. \label{qlimit}
\end{align}
This suggests a convenient parameterisation of the electric and magnetic charges by $\psi$ as
\begin{align}
 e=\sqrt{\frac{3b^2+1}{b^2-1}}\cos\psi,\quad g=\sqrt{\frac{3b^2+1}{b^2-1}}\sin\psi,
\end{align}
explicating the $U(1)$ symmetry of the Maxwell gauge field.

Taking $\epsilon\rightarrow 0$ for the metric, we find
\begin{align}
 \dif s^2&=\frac{-\ell^2(b^2-1)^2}{(x-b)^2}\sbrac{\frac{1}{q^2H^2(b-y_+)}\brac{-4u\dif\eta^2+\frac{\dif u^2}{u}}+\frac{\dif x^2}{G(x)}+G(x)\,\dif\phi^2}.
\end{align}
the $g_{\eta\eta}\dif\eta^2+g_{uu}\dif u^2$ part of the metric is conformal to a Ricci-flat spacetime in Rindler-type coordinates. We can transform it into a more familiar Minkowski form by
\begin{align}
 u=\frac{1}{4}q^2H^2(b-y_+)\brac{-\hat{\tau}^2+\hat{z}^2},\quad\tanh\eta=\frac{\hat{\tau}}{\hat{z}}, \label{RindlerTransform}
\end{align}
then the metric now becomes 
\begin{align}
 \dif s^2&=\frac{-\ell^2(b^2-1)^2}{(x-b)^2}\brac{-\dif\hat{\tau}+\dif\hat{z}^2+\frac{\dif x^2}{P(x)}+P(x)\,\dif\phi^2},
\end{align}
where
\begin{align}
 P(x)=\lim_{a\rightarrow b}\Pcal(x)=\frac{1-x^2}{b^2-1}\sbrac{(3b^2+1)(x-b)^2-\brac{b^2-1}\brac{2b(x-2b)+b^2+1}}
\end{align}
Next we do the same for the Maxwell potential. Taking Eq.~\Eqref{AdSCpot} and reparametrising with \Eqref{reparametrise}, followed by \Eqref{RindlerTransform}, we find, to leading order in $\epsilon$,
\begin{align}
 A=\frac{\mathrm{const}.}{\epsilon^2}+\sqrt{-\ell^2\brac{b^2-1}^2}\sqrt{\frac{3b^2+1}{b^2-1}}\sbrac{\half\cos\psi\brac{\hat{z}\dif\hat{\tau}-\hat{\tau}\dif\hat{z}}-x\sin\psi\dif\phi}.
\end{align}
The term containing $\epsilon^{-2}$ will diverge if we take the limit $\epsilon\rightarrow 0$. However, this term is constant and does not physically contribute to the electro-magnetic field. Therefore we shall ignore this term. Further invoking gauge invariance to remove a term proportional to $\hat{\tau}\dif\hat{z}$, the potential now becomes 
\begin{align}
 A=\sqrt{-\ell^2(b^2-1)^2(3b^2+1)}\sqrt{\frac{3b^2+1}{b^2-1}}\brac{\hat{z}\cos\psi\,\dif\hat{\tau}-x\sin\psi\,\dif\phi}.
\end{align}
Finally, let us simplify the solution a little further by rescaling,
\begin{align}
 \hat{\tau}\rightarrow\sqrt{b^2-1}\tau,\quad\hat{z}\rightarrow\sqrt{b^2-1}z,\quad\phi\rightarrow(b^2-1)\varphi,
\end{align}
along with a redefinition $G(x)\equiv \brac{b^2-1}P(x)$. The metric and gauge potential then become
\begin{subequations}\label{AdSMel}
\begin{align}
 \label{AdSMel_metric}
 \begin{split}
 \dif s^2&=\frac{-\ell^2(b^2-1)^3}{(x-b)^2}\brac{-\dif\tau^2+\dif z^2+\frac{\dif x^2}{G(x)}+G(x)\dif\varphi^2},\\
   G(x)&=\brac{1-x^2}\sbrac{\brac{3b^2+1}(x-b)^2-\brac{b^2-1}\brac{2b(x-2b)+b^2+1}},
 \end{split}\\
   A&=\sqrt{-\ell^2(b^2-1)^3(3b^2+1)}\brac{z\cos\psi\,\dif\tau-x\sin\psi\,\dif\varphi},\label{AdSMel_pot}
\end{align}
\end{subequations}
completing our derivation for the (A)dS Melvin spacetime.

It can be checked by direct calculation that \Eqref{AdSMel} is still a solution to the Einstein-Maxwell equations \Eqref{EME}. We also note that the coordinate ranges are slightly different for the case of positive and negative cosmological constants. In the former case, by Eq.~\Eqref{dSroots} we have $b<-1$. Therefore when $\ell^2<0$, the range of $x$ is 
\begin{align}
 -1< x < 1, \quad\mbox{(dS Melvin)}.
\end{align}
In this range, the metric has two poles at $x=\pm 1$ where $G(\pm1)=0$. Unlike the coordinate $\rho\in[0,\infty)$ in the $\Lambda=0$ Melvin spacetime, the coordinate $x\in[-1,1]$ of the dS Melvin solution is compact and should not be interpreted a `radius' in a cylindrically-symmetric spacetime. Hearkening back to the dS C-metric from which this solution is derived, $x$ might be more suitably regarded as the sine of a polar angle.

On the other hand, if the cosmological constant is negative, ($\ell^2>0$), the parameter $b$ lies in $-1<b<1$. By referring to the domain of the AdS C metric in Fig.~\ref{fig_DomainAdS}, our AdS Melvin limit is taken in the infintesimal neighbourhood of $y=b$, which is accessible only for 
\begin{align}
 b<x<1, \quad\mbox{(AdS Melvin)},
\end{align}
where $x=b$ is the conformal boundary of the spacetime. We shall hence take this as the coordinate range for the AdS Melvin spacetime. From this we conclude that this spacetime has one pole at $x=1$, where $G(1)=0$. Since $x$ is able to reach the conformal infinity in this case, it might be reasonable to interpret it as kind of `radial' coordinate. In fact, we shall show in the following section that this $x$ can be transformed into a form related to Poincar\'{e} coordinates of AdS.

\subsection{Derivation from the charged planar AdS black hole} \label{DeriveFromRN}

The magnetic AdS-Melvin spacetime with negative cosmological constant can also be derived from the charged AdS black hole with a planar horizon, which we shall refer to as the planar Reissner-Nordstr\"{o}m black hole in AdS, or the RN-AdS black hole. The RN-AdS metric and its gauge potential is given by \cite{Huang:1995zb}
\begin{subequations}
\begin{align}
 \dif s^2&=-f\dif t^2+f^{-1}\dif r^2+r^2\brac{\dif y^2+\dif z^2},\quad f=\frac{r^2}{\ell^2}-\frac{\mu}{r}+\frac{q^2}{r^2},\\
     A&=-\frac{q}{r}\,\dif t.
\end{align}
\end{subequations}
This is a solution to \Eqref{EME} describing a charged black hole with a planar horizon. Upon the following Wick rotations
\begin{align}
 t\rightarrow\im\tilde{\varphi},\quad y\rightarrow\im \tau,\quad q\rightarrow\im B, \label{TripleWick}
\end{align}
the solution becomes
\begin{subequations}
\begin{align}
 \dif s^2&=f\dif\tilde{\varphi}^2+\frac{\dif r^2}{f}+r^2\brac{-\dif\tau^2+\dif z^2},\quad f=\frac{r^2}{\ell^2}-\frac{\mu}{r}-\frac{B^2}{r^2},\label{AdS4soliton}\\
    A&=\frac{B}{r}\,\dif\tilde{\varphi}.
\end{align}
\end{subequations}
If we set $B=0$ above, we recover the familiar Horowitz-Myers AdS soliton \cite{Horowitz:1998ha} in four dimensions.

This solution actually our AdS Melvin spacetime \Eqref{AdSMel} in different coordinates. To see this, we apply the transformation
\begin{align}
 r\rightarrow\frac{\sqrt{-\ell^2(b^2-1)^3}}{x-b},\quad\tilde{\varphi}\rightarrow \sqrt{-\ell^2(b^2-1)^3}\varphi, \label{transform}
\end{align}
along with the identification
\begin{align}
 \mu=4\sqrt{-\ell^2(b^2-1)^3}b(b^2+1),\quad B=\sqrt{-\ell^2(b^2-1)^3(3b^2+1)},
\end{align}
we recover the magnetic AdS Melvin solution \Eqref{AdSMel}.

A well-known fact about the AdS soliton is that they are regular spacetimes which are asymptotically locally AdS. A conical singularity may potentially exist at $r=r_0$ where $f(r_0)=0$, but regularity can be ensured by appropriately fixing the periodicity of $\tilde{\varphi}$. The location $r=r_0$ is called the tip of the soliton. According to the transformation \Eqref{transform}, this tip corresponds to the root $x=1$ where $G(x=1)=0$.

Superficially, it may appear that the Wick-rotated RN-AdS solution has two free parameters $\mu$ and $B$, while the solution given in Eq.~\Eqref{AdSMel} is only parametrised by a single parameter $b$. Thus one might wonder if the latter is a special case of the former. However the two solutions are equivalent, since \Eqref{AdSMel} are in coordinates where the pole is fixed at $x=1$. On the other hand, the pole of the Wick-rotated RN-AdS is located at $r=r_0$ where $f(r_0)=0$, thus $r_0$ is determined in terms of $\mu$ and $B$ by the implicit equation
\begin{align}
 \mu=\frac{r_0^3}{\ell^2}-\frac{B^2}{r_0}. \label{r0}
\end{align}
One could, say, fix a coordinate system where $r_0=1$ so that the pole is now located at $r=1$. The the solution is then only parameterised by $B$, with $\mu=\ell^{-2}-B^2$ via Eq.~\Eqref{r0}.

\section{Some physical and geometrical properties} \label{Physical}

Let us explore some physical properties of the solution as found in Eq.~\Eqref{AdSMel}. As mentioned in the previous section, the coordinate range depends on the sign of the cosmological constant and thus determines the presence of conical singularities in our spacetime region of interest. For a positive cosmological constant ($\ell^2<0$), the  poles of the metric are located at $x=\pm1$, where $g_{\varphi\varphi}=0$. We check for the presence of conical singularities at $x=\pm1$ by calculating the circumference/radius ratios of infinitesimal circles at the poles. We find
\begin{align}
 \left.\frac{\mbox{circumference}}{\mbox{radius}}\right|_{x=\pm 1}=\kappa_\pm\Delta\varphi,
\end{align}
where
\begin{align}
 \kappa_\pm&=2(b\mp1)^2\brac{3b^2+1\pm 2b}.
\end{align}
When the cosmological constant is positive, where both $x=\pm1$ are accessible, one can eliminate the conical singularity at one of the poles by setting the periodicity of the coordinate $\varphi$ to be $\Delta\varphi=2\pi/\kappa_+$, or $\Delta\varphi=2\pi/\kappa_-$. Since we have only the freedom to fix $\Delta\varphi$, both conical singularities cannot be removed simultaneously. Thus the dS-Melvin universe inherits the cosmic strut or string that is responsible for the acceleration of the black hole.

On the other hand, when the cosmological constant is negative, the coordinate range is $b<x<+1$, where $x=b$ is the AdS boundary. Then the sole conical singularity can be removed by setting $\Delta\varphi=2\pi/\kappa_+$, leaving us with a regular metric in the Lorentzian region of the spacetime. Furthermore, close to the boundary at $x\sim b$, the metric is 
\begin{align}
 \dif s^2\sim\frac{\ell^2(1-b^2)^3}{(x-b)^2}\brac{-\dif\tau^2+\dif z^2+\frac{\dif x^2}{(1-b^2)^3}+(1-b^2)^3\,\dif\varphi^2}.
\end{align}
Up to straightforward linear transformation of the coordinates, this is the Poincar\'{e} section of AdS spacetime. Taking note that $\varphi$ is periodic, we conclude that the AdS Melvin solution is asymptotically locally AdS at $x\sim b$.

The Ricci scalar and Kretschmann invariant are
\begin{align}
 R&=-\frac{12}{\ell^2},\\
 R_{\mu\nu\sigma\lambda}R^{\mu\nu\sigma\lambda}&=\frac{1}{\ell^4\brac{b^2-1}^6}\big[56(3b^2+1)^2x^8-128b(9b^2+2)(3b^2+1)x^7\nonumber\\
  &\quad+32b^2(13+321b^4+138b^2)x^6-128b^3(9b^2+2)(15b^2+1)x^5\nonumber\\
  &\quad+80b^4(15b^2+1)^2x^4-128b^5(2-3b^2+93b^4)x^3+32b^6(13-30b^2+153b^4)x^2\nonumber\\
  &\quad-128b^7(9b^4-3b^2+2)x-192b^10+144b^12+416b^8+360b^4\nonumber\\
  &\quad-144b^2-480b^6\big].
\end{align}
So we can see that both are regular and finite everywhere within the coordinate range $-1<x<1$ or $b<x<1$.

The electric ($\Phi_E$) and magnetic ($\Phi_B$) flux across a constant $\tau$ and $z$ surface can be calculated by $\Phi_E=\int\dif\sigma^{\mu\nu} F_{\mu\nu}$ and $\Phi_B=\int\dif\sigma^{\mu\nu}\brac{*F}_{\mu\nu}$  where $\dif\sigma^{\mu\nu}$ is a area element of a 2-surface orthogonal to $\xi^\mu$ and $n^\nu$, which are respectively time-like and space-like unit Killing vectors along the $t$ and $z$ directions. Using $\dif\sigma_{\mu\nu}=-2\xi_{[\mu}n_{\nu]}\sqrt{\gamma}\dif^2y$, where $\gamma_{\mu\nu}$ is the induced metric on the 2-surface, we have
\begin{align}
 \Phi_E&=2\int\dif\varphi\dif x\,\sqrt{-\ell^2(b^2-1)^3(3b^2+1)}\cos\psi,\\
 \Phi_B&=2\int\dif\varphi\dif x\,\sqrt{-\ell^2(b^2-1)^3(3b^2+1)}\sin\psi.
\end{align}
In these coordinates, the flux per unit area is constant, lending favour to the notion that this spacetime carries a `homogeneous' electric/magnetic field along the $z$-direction, at least in this naive coordinate-dependent expression. An invariant  value of the fluxes can be obtained by integrating over $\varphi$ and $x$. Taking note that the range of $x$ depends on the sign of the cosmological constant, the total electric flux for positive and negative cosmological constants are
\begin{align}
 \mbox{dS, }\ell^2<0:\quad\Phi_E&=4\Delta\varphi\sqrt{-\ell^2(b^2-1)^3(3b^2+1)}\cos\psi,\\
 \mbox{AdS, }\ell^2>0:\quad\Phi_E&=2\Delta\varphi(1-b)\sqrt{-\ell^2(b^2-1)^3(3b^2+1)}\cos\psi.
\end{align}
Similarly, the magnetic flux for the two cases of the cosmological constant are
\begin{align}
 \mbox{dS, }\ell^2<0:\quad\Phi_B&=4\Delta\varphi\sqrt{-\ell^2(b^2-1)^3(3b^2+1)}\sin\psi,\\
 \mbox{AdS, }\ell^2>0:\quad\Phi_B&=2\Delta\varphi(1-b)\sqrt{-\ell^2(b^2-1)^3(3b^2+1)}\sin\psi.
\end{align}


\section{Other forms and limits} \label{OtherForms}

\subsection{Transformation to Astorino's form}

To recover the form of the (A)dS Melvin solution as presented by Astorino \cite{Astorino:2012zm}, we transform
\begin{align}
 x=b+\frac{\sqrt{-\ell^2(b^2-1)^3}}{1+\rho^2/4},\quad \varphi=\frac{2}{\sqrt{-\ell^2(b^2-1)^3}}\bar{\varphi},
\end{align}
and identifying 
\begin{align}
 B=\sqrt{-\ell^2(b^2-1)^3(3b^2+1)},\quad k=-16\sqrt{-\ell^2(b^2-1)^3}b(b^2+1),
\end{align}
the metric and gauge potential becomes 
\begin{subequations} \label{M_Astorino}
\begin{align}
 \dif s^2&=W(\rho)^2\brac{-\dif\tau^2+\dif z^2+\frac{\rho^2\dif\rho^2}{H(\rho)}}+\frac{H(\rho)^2}{W(\rho)^2}\dif\bar{\varphi}^2, \\
    W(\rho)&=1+\frac{\rho^2}{4},\quad H(\rho)=4B^2+kW(\rho)+\frac{4}{\ell^2}W(\rho)^4,\\
    A&=B\brac{z\cos\psi\,\dif\tau+\frac{2\sin\psi}{W(\rho)}\dif\bar{\varphi}}. \label{A_Astorino}
\end{align}
\end{subequations}
This is the form given by Astorino in \cite{Astorino:2012zm}. More precisely, Astornio's form is explicitly a magnetic universe, which corresponds to $\psi=\frac{\pi}{2}$ in Eq.~\Eqref{A_Astorino} above.

\subsection{The \texorpdfstring{$\Lambda=0$}{Lambda=0} limit}

We can recover the usual $\Lambda=0$ Melvin universe from \Eqref{AdSMel} by taking the limit
\begin{align}
 \ell^2\rightarrow\pm\infty,\quad b\rightarrow-1,\label{FlatLimit}
\end{align}
while keeping $\ell^2(b^2-1)^3$ finite. In terms of  Fig.~\ref{fig_Domains}, taking this limit is equivalent to reshaping the darker-shade box so that the top-left vertex touches the $x=y$ line, the result would be the $\Lambda=0$ diagram as appears in Fig.~\ref{fig_DomainFlat}. It should follow that the infinitesimal neighbourhood around $y=b$ shall turn into the $\Lambda=0$ Melvin limit of Harvdov\'{a} and Krtou\v{s} \cite{Havrdova:2006gi}.

To get this limit explicitly, let us first define 
\begin{align}
 -\ell^2\brac{b^2-1}^3=\frac{64}{B^2},
\end{align}
along with the rescalings
\begin{align}
 \tau\rightarrow\frac{B}{4}\tau,\quad z\rightarrow\frac{B}{4}z,\quad\varphi\rightarrow\frac{1}{16}\varphi.
\end{align}
Taking the limit \Eqref{FlatLimit}, we find
\begin{subequations}
\begin{align}
 \dif s^2&=\brac{1+\frac{B^2\rho^2}{4}}^2\brac{-\dif\tau^2+\dif z^2+\dif\rho^2}+\brac{1+\frac{B^2\rho^2}{4}}^{-2}\rho^2\,\dif\varphi^2,\\
     A&=Bz\cos\psi\,\dif\tau-\frac{B\rho^2\sin\psi}{2\brac{1+B^2\rho^2/4}}\,\dif\varphi,
\end{align}
\end{subequations}
which is the familiar Melvin universe with zero cosmological constant, particularly in the notation of \cite{Griffiths:2009dfa}.

\subsection{The Pleba\'{n}ski-Hacyan limit}

For the case of positive cosmological constant ($\ell^2<0$), we can also obtain the limit to the Pleba\'{n}ski-Hacyan \cite{PlebanskiHacyan,Griffiths:2009dfa} solution by first rescaling the coordinates by
\begin{align}
 \tau\rightarrow\frac{1}{\sqrt{6}b^2}\tau,\quad z\rightarrow\frac{1}{\sqrt{6}b^2}z,\quad\varphi\rightarrow\frac{1}{6b^4}\varphi,
\end{align}
and taking the limit $b\rightarrow-\infty$. The solution then becomes 
\begin{align}
 \dif s^2&=-\frac{\ell^2}{6}\brac{-\dif\tau^2+\dif z^2+\frac{\dif x^2}{1-x^2}+\brac{1-x^2}\dif\varphi^2},\\
       A&=\frac{\sqrt{-\ell^2}}{6}\brac{z\cos\psi\,\dif\tau+x\sin\psi\,\dif\varphi}.
\end{align}
This is an electro-vacuum, direct-product metric of the form $M_2\times S^2$ considered by Pleba\'{n}ski and Hacyan \cite{PlebanskiHacyan}.

\section{Geodesics} \label{Geodesics}

In this section we shall study the structure of the (A)dS Melvin spacetime via its gravitational effects on test particles within it, namely, its geodesics. The geodesics in the spacetime \Eqref{AdSMel} is described by a trajectory $x^\mu(\sigma)$, parametrised by $\sigma$. The trajectory is determined by the Lagrangian $\mathcal{L}=\half g_{\mu\nu}\dot{x}^\mu\dot{x}^\nu$, where over-dots denote derivatives with respect to $\sigma$. Given the metric \Eqref{AdSMel}, the Lagrangian is
\begin{align}
 \mathcal{L}&=\half\frac{\bar{H}}{(x-b)^2}\brac{-\dot{\tau}^2+\dot{z}^2+\frac{\dot{x}^2}{G}+G\dot{\varphi}^2}=\frac{\varepsilon}{2},
\end{align}
where $\bar{H}=-\ell^2(b^2-1)^3$. Due to the fact that the inner product of the four-velocities, $g_{\mu\nu}\dot{x}^\mu\dot{x}^\nu=\varepsilon$ should be constant along the trajectory. Using the freedom to rescale $\sigma$, the magnitude of $\varepsilon$ can be rescaled to unity if it is non-zero. Therefore we have 
\begin{align}
 \varepsilon=\left\{
    \begin{array}{rl}
     -1, & \mbox{for time-like geodesics},\\
      0, & \mbox{for null/light-like geodesics}.
    \end{array}
  \right.
\end{align}
Since the coordinates $t$, $z$, and $\varphi$ are cyclic in the Lagrangian, we have the first integrals
\begin{align}
 \dot{\tau}&=\frac{(x-b)^2}{\bar{H}}\mathcal{E},\\
 \dot{z}&=\frac{(x-b)^2}{\bar{H}}p,\\
 \dot{\varphi}&=\frac{(x-b)^2}{\bar{H}}\frac{L}{G},
\end{align}
where $\mathcal{E}$, $p$, and $L$ are conserved quantities which we may interpret as the particle's energy, linear $z$-momentum, and angular momentum, respectively. A second-order differential equation for $x$ can be obtained from the Euler-Lagrange equation $\frac{\dif}{\dif\sigma}\frac{\partial\mathcal{L}}{\partial\dot{x}}=\frac{\partial\mathcal{L}}{\partial x}$, though it is more convenient to use the invariance of $\varepsilon$ to obtain a first-order equation 
\begin{align}
 \frac{\bar{H}^2}{(x-b)^4}\frac{\dot{x}^2}{G}&=\mathcal{E}^2-V_{\mathrm{eff}}^2,
\end{align}
where $V_{\mathrm{eff}}^2$ is the effective potential
\begin{align}
 V_{\mathrm{eff}}^2&=p^2+\frac{L^2}{G}-\frac{\bar{H}\varepsilon}{(x-b)^2}.\label{V_eff}
\end{align}
From this effective potential, we can glean some qualitative aspects on the gravitational effects on a neutral test particle or photon in the (A)dS Melvin spacetime.

We start by considering null geodesics, where $\varepsilon=0$. In this case, we see that the only interesting term in Eq.~\Eqref{V_eff} is $L^2/G$. Similar to radial null geodesics in the $\Lambda=0$ Melvin universe, `radially' moving photons with $L=0$, the effective potential is a trivial constant and the motion is unbounded. On the other hand, if $L$ is non-zero, the photon encounters an infinite potential barrier when the denominator $G$ goes to zero, which is $x=\pm 1$ for the dS case and $x=b$ and $x=1$ for the AdS case. Furthermore, we can find `circular' orbits of constant $x$ by finding the minima of $V_{\mathrm{eff}}^2$. They are located at\footnote{Another trivial/uninteresting solution for $\dif(V_{\mathrm{eff}}^2)/\dif r=0$ is $x=b$.}
\begin{align}
 x=-\frac{2b}{3b^2+1}. \label{CircularPhotonOrbit}
\end{align}
This lies within the coordinate range $-1<x<1$ or $b<x<1$ as long as $b$ is negative. Therefore, for $b<0$, circular photon orbits are possible for any non-zero cosmological constant.

\begin{figure}
 \begin{center}
  \begin{subfigure}[b]{0.49\textwidth}
   \centering
   \includegraphics[scale=1]{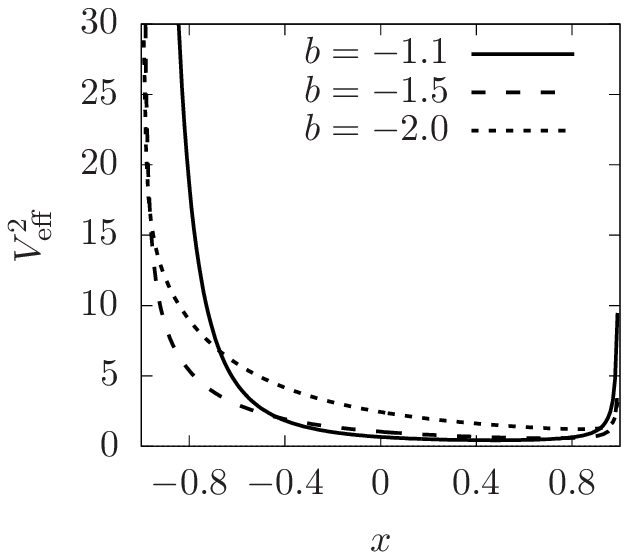}
   \caption{$\ell^2<0$ (dS)}
   \label{fig_VeffdS}
  \end{subfigure}
    \begin{subfigure}[b]{0.49\textwidth}
   \centering
   \includegraphics[scale=1]{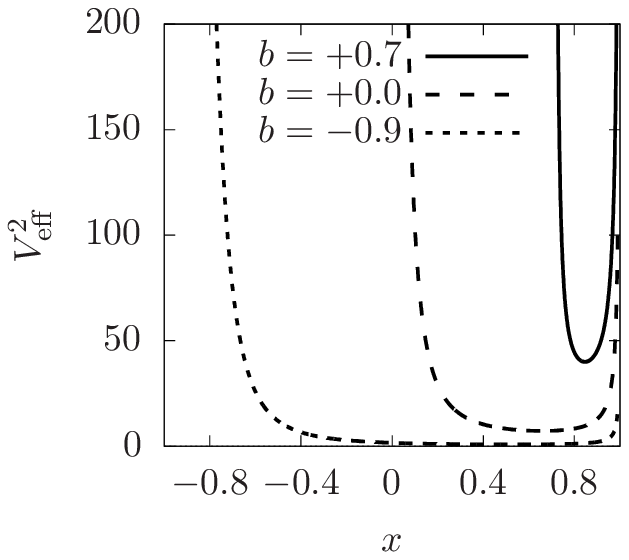}
   \caption{$\ell^2>0$ (AdS)}
   \label{fig_VeffAdS}
   \end{subfigure}
 \end{center}
 \caption{The plots of effective potentials for time-like particles with $p=0$, $L=2$ and various $b$ for the case of (a) positive cosmological constant and (b) negative cosmological constant.}
 \label{fig_Veff}
\end{figure}

For the case of time-like geodesics, there is now an additional term $\bar{H}\varepsilon/(x-b)^2$ where $\varepsilon=-1$. If the cosmological constant is positive, $b$ satisfies $b<1<x$ and this term does not diverge. If the angular momentum of the particle is zero, the effective potential is finite everywhere. So infinite potential barriers at $x=\pm 1$ only occur when angular momentum is non-zero. Therefore such time-like particles are in a potential well between $-1<x<1$. This potential has a minima, though calculation of $\dif\brac{V_{\mathrm{eff}}^2}/\dif x$ results in a cumbersome expression. Nevertheless the shapes of the potentials are shown in Fig.~\ref{fig_VeffdS}. For some fixed energy $\mathcal{E}$, a particle may oscillate about $x_1\leq x\leq x_2$ where $x_{1,2}$ are solutions to the equation $\mathcal{E}^2=V_{\mathrm{eff}}^2$. If $\mathcal{E}$ coincides with the minima of $V_{\mathrm{eff}}^2$, the particle would execute a `circular motion' at constant $x$.

If the cosmological constant is negative, the range of $b$ is $-1<b<1$. Therefore the term $\bar{H}\varepsilon/(x-b)^2$ is a potential barrier when the particle approaches the conformal infinity $x=b$. From Fig.~\ref{fig_VeffAdS}, we see that within $b<x<1$, the effective potential gets wider and and deeper as $b$ is decreased from $b=1$ towards $b=-1$. Accordingly, particles of energy $\mathcal{E}$ would oscillate between the roots of $\mathcal{E}^2=V_{\mathrm{eff}}^2$, and will be in constant-$x$ motion if the energy coincides with the minima of the effective potential. Unlike the dS case, the infinite barrier remains present for $L=0$. This is reminiscent to the negative cosmological constant of AdS acting as a confining box, keeping the particle from escaping to infinity.

\section{Higher-dimensional generalisation} \label{HigherD}

Throughout this paper, we have so far considered the (A)dS Melvin solution specifically in $D=4$ dimensional Einstein-Maxwell gravity. Aside from the obvious reason that this has the most observational/experimental relevance, we have been constrained to this dimensionality because of its derivation from the (A)dS C-metric, which is a solution only known for $D=4$.\footnote{To the author's knowledge, possible approaches to extending the C-metric to higher dimensions have been discussed \cite{Charmousis:2003wm,Kubiznak:2007kh,Kodama:2008wf}, though an exact solution remains to be seen.} 

The higher-dimensional magnetised spacetimes with $\Lambda=0$ are readily obtained since the Harrison transformation is still applicable in $D\geq 4$. Indeed, various magnetised spacetimes with and without black holes have been found \cite{Ortaggio:2004kr}. In fact, further generalisations into Einstein-Maxwell-dilaton gravity still hold \cite{Galtsov:1998mhf,Yazadjiev:2005gs,Lim:2017dqw}. For $\Lambda\neq0$, we do not have the Harrison transformation, and thus one has to find alternative methods to obtain the (A)dS Melvin solution.

For the case of a negative cosmological constant, an obvious possibility comes to mind: We have shown in Sec.~\ref{DeriveFromRN} how to derive the AdS Melvin universe starting from the planar Reissner-Nordstr\"{o}m-AdS solution, which is readily extendible to higher dimensions. For dimensions $D\geq 4$, the metric and gauge potential is given by
\begin{subequations}\label{RNAdS}
\begin{align}
 \dif s^2&=-f\dif t^2+f^{-1}\dif r^2+r^2\brac{\dif y^2+\dif x_2^2+\ldots+\dif x_{D-2}^2},\quad f=\frac{r^2}{\ell^2}-\frac{\mu}{r^{D-3}}+\frac{q^2}{r^{2(D-3)}},\\
       A&=-\sqrt{\half(D-2)(D-3)}\frac{q}{r^{D-3}}\dif t.
\end{align}
\end{subequations}
It is a solution to the $D$-dimensional Einstein-Maxwell equations
\begin{align}
 R_{\mu\nu}&=\frac{2\Lambda}{D-2}g_{\mu\nu}+2F_{\mu\lambda}{F_\nu}^\lambda-\frac{1}{D-2}F^2g_{\mu\nu},\\
 \nabla_\lambda F^{\lambda\nu}&=0.
\end{align}
Under the Wick rotations \Eqref{TripleWick}, the solution \Eqref{RNAdS} becomes
\begin{subequations}\label{RNAdS_soliton}
\begin{align}
 \dif s^2&=f\dif\tilde{\varphi}^2+f^{-1}\dif r^2+r^2\eta_{ab}\dif y^a\dif y^b,\quad f=\frac{r^2}{\ell^2}-\frac{\mu}{r^{D-3}}-\frac{B^2}{r^{2(D-3)}},\label{RNAdS_soliton_metric}\\
       A&=\sqrt{\half(D-2)(D-3)}\frac{B}{r^{D-3}}\dif\tilde{\varphi},\label{RNAdS_soliton_pot}
\end{align}
\end{subequations}
where $\eta_{ab}\dif y^a\dif y^b=-\dif\tau^2+\dif x_2^2+\ldots+\dif x_{D-2}^2$, which is a $(D-2)$-dimensional Minkowski spacetime.

Note that we can replace $\eta_{ab}$ in \Eqref{RNAdS_soliton_metric} with another Ricci-flat metric and the result would still solve the Einstein-Maxwell equations. For $D>5$, this suggests a tempting possibility to replace it with a Schwarzschild metric, thus making it a type of black hole solution. This would be a subject for another study beyond the scope of this paper, though we shall comment briefly on this prospect in Sec.~\ref{Discussion}.

\section{Conclusion} \label{Discussion}

By taking an appropriate limit of the (A)dS C-metric, we have extracted solution describing a electric/magnetic universe which solves the Einstein-Maxwell equations with a cosmological constant. This is a generalisation of Havrdov\'{a} and Krtou\v{s}'s procedure for the derivation of the Melvin universe from the $\Lambda=0$ C-metric, and supplements Astorino's solution which was derived from  generalised version of the Ernst formalism.

From the analysis of the electric/magnetic flux, we see that this solution indeed carries a homogeneous electric/magnetic fields. If the cosmological constant is positive, the spacetime necessarily carries a conical singularity at either one of the poles $x=\pm 1$, depending on how we fix the periodicity of the azimuthal coordinate $\varphi$. On the other hand, if the cosmological constant is negative, the spacetime only has one pole, and fixing $\Delta\varphi$ appropriately removes the sole conical singularity at $x=+1$.

We have also studied some qualitative features of null and time-like geodesics in the (A)dS Melvin universe. Similar to the $\Lambda=0$ case, particles with non-zero angular momentum would oscillate about the minimum of an effective potential well. Circular null and time-like orbits are possible if the energy is fixed to be equal to the minima of the effective potential.


An issue that naturally springs into mind is whether it is possible to add a black hole into the (A)dS Melvin solution, which would be a generalisation of the Ernst spacetime \cite{Ernst:1975} to include a cosmological constant. The Ernst spacetime was obtained by applying the Harrison transformation to the Schwarzschild black hole, which is a procedure not available in the presence of a cosmological constant. Instead, one could attempt to have an electrified/magnetised (A)dS black hole by solving the Einstein-Maxwell equations directly, though we naturally expect the equations of motion to be complicated. 

Meanwhile, we note that in the case of negative cosmological constant, Sec.~\ref{Derivation} and \ref{HigherD} tells us that the AdS Melvin spacetime can be viewed as a charged form of an AdS soliton. 
As alluded to in Sec.~\ref{HigherD}, one could replace $\eta_{ab}$ in Eq.~\Eqref{RNAdS_soliton_metric} with any Ricci-flat metric and still solve the Einstein-Maxwell equations. So one could easily replace it it with a $(D-2)$-dimensional Schwarzschild metric, modifying \Eqref{RNAdS_soliton_metric} to 
\begin{align}
 \dif s^2&=f\dif\tilde{\varphi}^2+f^{-1}\dif r^2+r^2\brac{-V\dif\tau^2+V^{-1}\dif R^2+R^2\dif\Omega^2_{(D-4)}}, \label{NotErnst}
\end{align}
where $V=1-R_+^{D-5}/R^{D-5}$. The solution would indeed be of a black-hole type with a horizon at $R=R_+$. Though we probably should not view \Eqref{NotErnst} as a $\Lambda\neq0$ analogue of the Ernst solution, which describes an isolated black hole with a compact horizon immersed in an external magnetic field. The horizon in \Eqref{NotErnst} is clearly not compact, as it is smeared along the non-compact $r$ direction. For the case of vanishing electro-magnetic fields, Eq.~\Eqref{NotErnst} and similar solutions are known as the AdS soliton \emph{string} and has been studied by Haehl in \cite{Haehl:2012tw}. In light of this, a proper generalisation of the Ernst metric describing an isolated (A)dS black hole immersed in an electric/magnetic field would be an interesting avenue for further study.

\section*{Acknowledgements}
The author thanks Zarmina Zaman Babar for discussions during the initial stages of this work.

\bibliographystyle{adsmel}

\bibliography{adsmel}

\end{document}